\documentclass{article}
\usepackage{epsfig}
\usepackage{amssymb}
\begin{document}
\begin{titlepage}
\begin{center}
\vspace{3.0cm}


{\Large {\bf  Black Holes {\em versus} Strange Quark Matter}} 

\vspace{2.5cm}
{\large Ewa
G\l{}adysz-Dziadu\'s}
 
\vspace{0.6cm}{\it
  Institute of Nuclear Physics, PAN,  Cracow, Poland}\\
\vspace*{8mm}

 February 2005\\
\end{center}
\vspace{0.4cm}
\end{titlepage}
\newpage 
\newpage

\section{Introduction}
 The interpretation  of Centauro-like events, observed by 
cosmic-ray 
mountain experiments
\cite{Hasegawa,Gladysz_review} still remains the open question.
A lot of models  proposed to explain Centauros and other
 related phenomena have been described and discussed in 
\cite{Gladysz_review}. To this list the new ideas based on DCC 
mechanism 
\cite{Silagadze, Serreau} or mini black holes 
evaporation \cite{Longo, Mironov} should be added. The last ones are  
based on the scenarios with large extra dimensions in which the 
fundamental Planck scale could be in TeV range.
 One of the most 
exciting consequences of a low fundamental Planck scale is the 
possibility of producing black holes and observing them in future 
colliders or in cosmic ray experiments \cite{Cavaglia_rev}. 
 The purpose  
of this work is to compare the strange quark matter and mini 
black hole scenarios and in particular to give some comments to the 
paper 
\cite{Mironov}. 

\section{Are Black Holes the best explanation of the Centauro 
events?}

If gravity propagates in $d+4$ dimensions while the other fields are 
confined to a 3-brane, the 4-dimensional Planck scale is given by 
$M^{2}_{pl} = M^{n+2}_{P}V_{n} = G^{-1}_{n+4}V_{n}$, where $M_{P}$ 
and $G^{-1}_{n+4}$ are respectively   
the fundamental Planck scale and the gravity constant in $4+n$ dimensions 
and $V_{n} = (2\pi 
R)^{n}$ is the volume of the n-torus of the compact space \cite{MBH_LHC}. 
In 
models with  large extra dimensions $R$ the fundamental Planck 
scale can be of the order of 1 TeV and the black holes could be 
produced at LHC energies. It is expected that these mini black holes 
(MBH) will decay very 
rapidily, with a lifetime $\tau \sim 1/M_{P} \sim 10^{-27}$ s for 
$M_{P} \simeq$ 1 TeV, in several stages. The  semiclassical Hawking 
evaporation
 will provide a high multiplicity of particles~\footnote{ 
 In some scenarios the expected multiplicity from black holes decay 
is quite 
 low and it can be futher reduced by quantum 
gravity effects \cite{Cavaglia_mult}, 
even 
at LHC energies is only $\sim$ 6 \cite{Longo}.} and 
a characteristic black-body type spectrum. Emission of all degrees of 
fredom are equally probable. It is expected that after evaporation they 
can leave 
remnants which are some excited stringy states.
A. Mironov, A. Morozov and T.N. Tomaras \cite{Mironov}  argue that 
the signals 
expected from the evaporation of mini black holes, predicted by 
TeV-Gravity models,  are similar to the 
characteristics of the Centauro events \cite{Mironov}. Morever, they
 suggest that this interpretation is better, or at least not worse 
than any other proposed so far because "its advantage is the 
potential to explain all the CLE \footnote{Centauro like events.}
properties at once (no worse than the other approaches, targeted at 
these properties) and, furthermore it offers somewhat less freedom to 
arbitrarily adjust different predictions".

We agree that the evaporating mini black holes are good candidates 
for the origin of unusual cosmic-ray events, however, to draw the 
 above cited conclusion, 
this scenario needs much more detail studies and answering some 
important questions. The detailed calculations and simulations
of the production and decay of mini black holes are necessary  
 to compare the characteristics of simulated events with 
experimentally observed Centauros as well as to design the 
  future accelerator experiments to be sensitive to such objects
  production and detection. It is true that more recently the 
simulations of cascades, generated  by the decay of mini black holes 
formed in the ultra high energy collisions of the cosmic ray 
particles with the 
atmospheric nuclei have been done \cite{MBH_2}. The simulations have 
been performed for several values of the large extra dimensions and 
for a variety of altitudes of the initial interactions, and compared 
with the standard events. However, to settle the Centauro question, 
much more detailed work, and especially consideration of the 
signals from the black holes produced rather close to the detector 
(at heights smaller that $\sim$ 1000 m above the chamber) should be 
done.
 
 Majority 
 of the proposed models, as it was stressed 
in 
\cite{Gladysz_review}, are not able to explain simultaneously all 
features of 
the Centauro-like events and they are mainly concentrated on the
interpretation of the basic Centauro anomaly, i.e. the enormous
hadron-rich composition. The exception are, however,
the Strange Quark Matter (SQM)  based scenarios, such as the model 
 proposed 
and 
decribed in \cite{SQM_model_1} and supplemented with the idea of  
formation  and passage of strangelets through the matter 
\cite{SQM_model_2,str_cosmic}.
 At this 
stage of investigation, it  offers more 
 complete and quantitive description of the cosmic ray exotic 
phenomena than other approaches, among them also the mini black holes 
scenario. However, the mini-black hole concept has also some 
interesting points and it is worth of further considerations.
 The common advantage of both scenarios is the 
 expectation of the threshold energy necessary for generation 
of Centauro-like phenomena. Let us  try to compare and 
discuss 
some  essential  points.

\subsection{Multiplicities and hadron rich composition}

The main, widely known feature of the Centauro-like events is their
anomalous hadronic to photonic ratio and additionally rather small 
observed multiplicity, as compared to that expected from 
nucleus-nucleus collisions at the same energy range.

The hadron rich composition of the events can be easily understood 
by assuming the existence of the strange quark matter. Such state of 
 matter, containing a comparable fraction of $u$, $d$ and $s$ quarks, 
  initially  proposed by E. Witten 
\cite{Witten} is  recently the subject of many theoretical 
works. The main argument for the stability of such system is the 
introduction of a third flavour what creates an additional 
Fermi well and thus reduces the energy relative to a two-flavor 
system. It should be stressed that a lot of cosmic ray anomalies, 
 such 
as massive and relatively low charged objects 
\cite{Saito,Price,Ichimura}, muon bundles from CosmoLEP 
\cite{Cosmolep}, delayed neutrons observed with EAS instalation
``Hadron'' \cite{Zhdanov}, Centauro related phenomena 
\cite{SQM_model_1,SQM_model_2,
str_cosmic,Shaulov,Wlodarczyk_str} are proposed to be explained by 
assuming 
the presence of strangelets 
in cosmic ray spectrum \cite{Wlodarczyk_rev}. To this list the 
 seismic events with the properties of the passage 
of SQM droplets 
through the Earth \cite{seismic} and the candidate for a metastable 
strangelet 
found with the AMS detector \cite{AMS_strangelet} should be added.
    
According to the model developed in \cite{SQM_model_1} Centauro arises 
through the hadronization of a SQM fireball, produced in 
a nucleus-nucleus collision in the upper atmosphere. The fireball
formed in the forward rapidity and high baryon density environment
 initially consists of $u$ and $d$ quarks and gluons only.
 The high baryon chemical potential inhibits the creation of
$u \overline{u}$ and $d \overline{d}$ quark pairs, resulting in the 
fragmentation of gluons predominantly into $s \overline{s}$  pairs.
This leads to the strong suppression of pions and hence of photons
in the process of its hadronization.
To form a pion, q and $\overline q$ must exist. Large baryon chemical
potential $\mu_{b} = 3\mu_{q}$  suppresses the production of $\overline u$ 
and 
$\overline d$
quarks because
\begin{equation}
 N_{\overline q} \propto exp(-\mu_{q}/T).
\end{equation}
 In
addition, the production of quarks is strongly prohibited      
due to the Pauli blocking.

 Thus, the number of pions, formed by
created $q\overline q$ pairs, is proportional to 
\begin{equation}
N_{pions}\propto exp(-(4\mu_{q}+m_{\pi})/T),
\end{equation}
and the pion--to--nucleon ratio is proportional to 
\begin{equation}
\frac{N_{pions}}{N_{nucl.}}\propto \frac{3}{2}{[-\frac{4}{3}\mu_{b}
-m_{\pi}+M_{n}]/T}\simeq 7\times 10^{-6}.
\end{equation}
 Such estimate has been obtained in \cite{SQM_model_1} by
assuming that nucleons are formed by existing {\it u,d} quarks.  
 The  values of thermodynamical 
quantities, such as the energy density
$\varepsilon = 2.4$ GeV/fm$^{3}$, the quark chemical potential $\mu_{q} = 
600$ 
MeV and the temperature $T$ = 130 MeV 
are not free model parameters but they have 
been extracted  
from analysis of experimental characteristics  available 
 for the five ``classical'' Chacaltaya Centauros. So,  in the SQM 
fireball 
picture  the 
strong 
suppression of the electromagnetic component is the natural 
consequence of the nucleus-nucleus collisions at energies 
high enough to produce, at the forward rapidity region, the fireball of 
the 
high baryon density quark gluon plasma. The mechanism of the 
fireball formation proposed in
\cite{SQM_model_1,SQM_model_2}  
 does not need any additional assumptions and 
fitted parameters and successfully (and quantitatively)  explains the 
features of
 Centauros detected at mountain top cosmic ray experiments  (i.e. hadron 
multiplicity $N_{h} \sim 75$, suppression of 
electromagnetic component, shape of a pseudorapidity 
distribution).
The Monte Carlo generator of Centauro events, CNGEN, constructed on the 
basis of this 
model not only satistactorily describes the main features of cosmic ray 
Centauros but also gives quantative predictions concerning these events 
possibly produced at LHC accelerator \cite{Gladysz_review,Centauro_sim}.

The authors of \cite{Mironov} suggest that the strong suppression of 
electromagnetic component observed in Centauro events could be  
the result of evaporation of  MBHs. In this process all kinds of 
particles of the Standard Model can be produced, so the resulting
photon/hadron ratio should be lower than for ordinary hadronic 
interactions. According to \cite{Cavaglia_rev,Cavaglia_mult} the 
hadron to lepton 
ratio is about 5:1, what leads to a  hadronic to photonic ratio of 
about 100:1. No constraint on isospin in the final state of 
MBH 
 is  the additional argument, given in 
\cite{Longo}, for 
possibility of appearance of large fluctuations in the ratio of 
charged to neutral pions. However, at the present level of the 
study, above mentioned arguments have strictly qualitative 
character.  It was not shown  for what values of the parameters 
the MBH scenario could quantitatevely describe the observed 
characteristic features of Centauros, such as for example the observed 
multiplicity of 
 electromagnetic and hadronic component,  and in particular if the 
 understanding of the extreme situation, the so called clean 
Centauros (with the totally 
reduced electromagnetic component) could be  possible.
 According to \cite{Mironov} the number of emitted particles 
$N_{in} \simeq M_{BH}/2T_{BH}$ (where $M_{BH}$  and $T_{BH}$ are 
the mass and temperature of the black hole respectively)  and  the 
multiplicity of the particles 
observed $N_{obs}$ should be calculated using fragmentation functions 
and will strongly depend on the BH parameters.

\subsection{Transverse momentum of produced secondaries}

Transverse momentum $p_{T}$ of particles produced in the 
Centauro-like events 
is also surprising. The experimental data indicate that we 
observe two components of the transverse momentum.

\begin{itemize}
\item  The first one
constitutes the particles with high transverse momentum, estimated to 
be $\langle p_{T} \rangle \sim$ 1.7 GeV/c for Centauros and
$\langle p_{T} \rangle \sim$ 10-15 GeV/c for Chirons. These values have 
been obtained from the formula:
\begin{equation} 
 p_{T} \simeq R\cdot K_{\gamma}\cdot E/H
\end{equation}
 where R is 
the distance of the particle from the family centre, 
$K_{\gamma}$ is the inelasticity coefficient, E is the energy of 
the electromagnetic shower produced in the apparatus, and  H is the 
height of the interaction point.

 We should remember, however, that 
 estimated in this way values of the transverse 
momenta were obtained by assuming the gamma inelasticity 
coefficient $K_{\gamma} \simeq 0.2$ and  they suffer from 
 some experimental 
ambiguities. First of all   the 
measurement of 
the 
height
of the interaction point above the chamber is the 
serious problem \cite{Gladysz_review}. 
 The heights have been measured directly, through the 
 shower geometry, only for a few exotic events found in 
Chacaltaya chambers (for three Centauros, two Mini-Centauros and 
three
Chirons). The transverse momenta of secondaries in the other events 
have been estimated in a more speculative way.
\item
 The second component is seen in the narrowly 
collimated 
 jets of 
particles (called mini-clusters) firstly observed in Chirons,
 later also in other Centauro-type events. The ratio of $\langle 
E( \gamma )R \rangle$ of hadron cascades in Centauro-like events to 
the $\langle E(\gamma)R \rangle$ in mini-clusters has been found to be 
surprisingly large ($\sim 300$) and their intrinsic $p_{T}$ is supposed to 
be  
extremely small ($\sim 10-20$ MeV/c). In addition, the mini-clusters 
are characterized by very strong penetrative power in the matter
(see the next subsection).
\end{itemize}
{\it In the SQM model \cite{SQM_model_1,SQM_model_2, str_cosmic} the high 
transverse 
momentum of 
secondaries can be the result of the explosive decay of the Centauro 
fireball and the highly collimated clusters of cascades could be  a 
natural 
consequence of the evaporation of cold strangelets, during their passage 
through the apparatus matter.} In 
\cite{Gladysz_review,Centauro_sim,str_mixed} we have presented the 
results of 
the 
simulations done by means of our Centauro event generator, CNGEN,
 and in particular, we have shown  
that the average transverse momentum of the Centauro fireball decay 
products is several times higher than that for normal events. 
For example the average transverse momentum of particles coming 
from the decay of the Centauro fireball characterized by a 
temperature T = 130 MeV \footnote{T = 130 MeV is  estimated from 
characteristics of the 
cosmic ray Centauros.} and produced in $PbPb$ collisions at LHC 
energies ($\sqrt{s}$ = 5.5 A$\cdot$TeV) is 1.6 GeV/c, in comparison 
with 
$p_{T} \simeq 0.44$ GeV/c obtained for conventional events (HIJING 
generator).

 The question of the existence and the understanding of the  
exotic features of mini-clusters, and in particular, their 
surprisingly small lateral spread is not considered 
by 
authors of \cite{Mironov} in the MBH scenario. {\it It is not clear what 
mechanism connected with the MBH phenomenon could be 
responsible
for the production of very narrow and strongly penetrating jets
of particles?}

On the other hand {\it the appearance of the particles with high 
transverse momenta
 is strongly stressed by authors of MHB scenarios. They 
predict 
that the decay of  MBHs looks like a typical fireball situation, but 
its temperature is very high, $T_{BH} \sim$ 1 TeV.} The 
simulations done in \cite{MBH_LHC} for the $pp$ and $PbPb$ collisions 
at the LHC energies
 showed that  hadrons from mini black hole 
decay dominate the background at transverse momenta $p_{T} \gtrsim$ 30 - 
100 GeV/c. The consequences of the production of the particles 
having such enormously large transverse momenta on the 
characteristics of the Centauro-like events should be very pronounced. At first,  
only a fraction of the particles radiated by a black hole could be 
observed in the chambers. The substantial part of them miss 
 the 
detector. For example the low energy ($\sim$ 2 TeV) particle 
\footnote{The energy 
threshold 
for the X-ray films is $\sim$ 2 TeV. } emitted with $p_{T} \simeq$ 1 
TeV/c
at the height of 1000 m above the detector, reaching  the level of the 
chamber 
will be distanced at   $R \simeq$ 
58 m from the family centre. Keeping in 
mind that the areas of  
typical emulsion chambers used in these experiments  are of the order 
 of tens $m^{2}$ and additionally that the area, scanned for  finding 
 the
cascades belonging to the same event is usually limited to the 
 circle of a radius $ R \simeq$ 15 cm \footnote{In the part of the 
experimental material,
measured 
by our Cracow group, this area has been increased to  $R \simeq$ 30 cm.}
such high $p_{T}$ particles must escape the detection. Only very  
high
energy particles and/or these produced at low height above the chamber
have a chance to be seen in the detector. Table~1
illustrates this question.  

\begin{table}[]
\label{tab:pT}  
\caption{Distances $R$ of the high $p_{T}$ particles from the centre of 
the 
family at the detector level }
\begin{center}
\vspace{0.3cm}
\begin{tabular}{|c||ccc|}
\hline
 & & &  \\
 R [m]& E [TeV] & H [m] & $p_{T}$ [GeV/c] \\
 &&&\\
\hline
\hline
&&&\\
3&1&100&30\\
58&2&100&1000\\
30&1&1000&30\\
580&2&1000&1000\\
{\bf 0.3}&{\bf 10}&{\bf 100}&{\bf 30}\\
10&10&100&1000\\
{\bf 0.03}&{\bf 100}&{\bf 100}&{\bf 30}\\
{\bf 0.1}&{\bf 100}&{\bf 100}&{ \bf 100}\\
1&100&100&1000\\
 {\bf 0.3}&{\bf 100}&{\bf 1000}& {\bf 30}\\
 1&100&1000&100\\
 &&&\\
\hline
\end{tabular}
\end{center} 
\end{table}

 Thus,  the decaying MBHs will produce events, characterized by a  very 
wide lateral spread, much 
wider than that conventionally limited to the circle of a radius $R$ = 15 
cm in the 
standard procedure of measurements. The simulations of the cascades 
through the atmosphere done in \cite{MBH_2} also support this 
statement, claiming that the wide lateral distribution will be the 
most striking signature of a black hole event.

 The second important point is that the production angles $\theta_{prod}$ 
of such 
extremely high $p_{T}$ particles are enough big to apparently change the
direction defined by the projectile arrival. In the typical ``normal'' 
events,
 with $p_{T} \sim$ 0.4 GeV/c the corresponding $\theta_{prod}$ is $\sim
2\cdot10^{-4}$ rad $\simeq 0.01^{o}$ for $\sim$ 2 TeV particles, and
smaller for higher energy particles. Thus, the experimental procedure 
used to identify  particles belonging to the same event is 
based on the requirement of the same flight direction for all members 
of the family. The experimental uncertainties in the measuring of the 
zenithal $\theta$  and azimuthal $\phi$  angles are much bigger than 
the change in the 
particle direction caused by its production angle. As the example, 
for the X-ray films 
technique used in the Pamir Experiment, the estimated error in 
the solid angle $\Delta \Omega \simeq 0.01$ sr \cite{Pamir_Baradzei}. 
In the typical 
procedure of elaboration of the events, all the showers with angles the 
same  within the errors (in the Pamir 
  $\Delta \theta \simeq 3^{0}$ and $\Delta \phi \simeq 30^{0}$ 
 ($\Delta \theta \simeq 6^{0}$ and $\Delta \phi \simeq 60^{0}$) for 
photon  
 (hadron) cascades
 are considered 
to be the members of the same family. 

The situation is different in the case of huge $p_{T}$.
 The particles of energy 2 TeV (10 TeV) and $p_{T} 
\sim$ 1 TeV, generated in the evaporation of MBHs, will be emitted at
$\theta_{prod} \sim$ 0.52 rad $\simeq 30^{0}$ ($\theta_{prod} \sim$ 
0.1 
rad 
$\simeq 5.7^{0}$). Such angles
 are greater than the values of experimental errors imposed
on the measured angles. In consequence,  the very high 
$p_{T}$ particles could not be recognised as  members of the same 
family.

 At the moment, we should invoke the recent papers 
\cite{CenI_Tamada, CenI_Kopenkin_ICRC, CenI_Kopenkin}, in which 
authors describe the 
re-examined Centauro I event. Centauro I was the first and the most 
spectacular  event of this 
type. It was found, in 1972, in 
the 
two-storey Chacaltaya chamber No.15. According to the 
original analysis \cite{Hasegawa},
the event consisted of two groups of showers: one group of 7 showers was 
observed in the lower 
chamber (called in \cite{CenI_Tamada} {\it S55} family) and the other one 
of 43 showers \footnote{ 137 showers with detection threshold of 
0.1 TeV detected in emulsion plates \cite{CenI_Tamada}.}(called in 
\cite{CenI_Tamada} {\it I12} family) was found in the upper chamber. 
The authors of  
the new analysis \cite{CenI_Tamada, CenI_Kopenkin_ICRC,CenI_Kopenkin} claimed that they 
discovered 
 differences in angles between the  cascades registered in 
the upper  and  lower blocks. 
They found \cite{CenI_Kopenkin_ICRC, CenI_Kopenkin} that $tg 
\theta \sim 0.3 \pm 0.1$, 
$\phi \sim (130 \pm 10)^{0}$
 and $tg \theta \simeq 0.4 \pm 0.1$, $\phi \sim (90 \pm 10)^{0}$ for 
cascades in the lower and upper 
chambers 
respectively (according to \cite{CenI_Tamada}  $tg \theta = 
0.3 \pm 0.1$, 
$\phi \sim (130 \pm 10)^{0}$
 and $tg \theta \simeq 0.3 \pm 0.1$, $\phi \sim (90 \pm 10)^{0}$ for 
cascades in the lower and upper 
chambers 
respectively). In connection with it, the authors of these papers 
assume that these 
groups of showers do 
not 
consitute the same family but they come from two different primary 
projectiles, accidentally flying in the  very close directions. The further conclusion from \cite{CenI_Kopenkin_ICRC} is that in the case of the ideal detector, the SQM scenario could be the most plausible explanation, although the authors suggest the Chacaltaya detector problems  as the most likely cause of such puzzle event.
Then, according to \cite{CenI_Kopenkin}  
  the new 
physics is not needed for Centauro I explanation, and the event 
observed in the lower chamber could be the air family passing through
a gap between blocks in the upper detector.
 In such case, however, immediately the question arises about 
the evidence
 on anti-Centauro events,
 artificially produced in the possible gaps between blocks 
 in the lower chamber. The 
considerations done in 
 \cite{CenI_Tamada} lead the authors   to the opinion 
that the Centauro I passed the upper detector with leaving no (or a 
single) shower in the upper detector and 
 the event seems to be  even  much more surprising 
than before. They suggest that the event could be
 explained by SQM scenario, assuming that large quark globs,
 present in the primary cosmic ray spectrum,
 fragment in the atmosphere  in smaller quarks lumps and
 they hit the emulsion chamber.

 It seems to us that other  possibility of the understanding of 
this puzzle is to agree with the original analysis of the Centauro I,
 by allowing the larger errors in the shower direction and assuming 
the same 
origin of 
the {\it S55} and {\it I12} groups. The 
 admission  of  larger errors on the measured azimuthal and 
zenithal angles is the straightforward consequence of assuming  
the existence of phenomena in which very high 
$p_{T}$ 
particles are generated,  such as MBHs scenario or generally the explosive 
decay of the 
 large mass fireballs (e.g. the SQM  fireball with high T 
and/or $\mu$).

It should be stressed that Centauro I is not the sole event with such 
peculiar characteristics. The other one is the Centauro-like event
found more recently in Chacaltaya two-storey chamber no. 22.
According to descriptions done in 
\cite{Cen_new}  no showers in the upper chamber have been 
observed and although some cascades can be missed in the gaps 
between blocks, the authors think that it does not seem  possible 
that gaps 
 between blocks could  be 
the 
reason of missing all showers in the family. 
 May be the particles with very high $p_{T}$  have been 
lost?

The phenomenon, which looks like  the  appearance of   
two familes with very close arrival directions has been also observed by 
our
 Cracow group. In particular, doing the final analysis of the 
Centauro-like 
event
 found in the thick Pb emulsion chamber of the Pamir 
Experiment \cite{Cen_Krakow} we have excluded from the event 30 showers
\footnote{After excluding 30 cascades with slightly different angles,
the multiplicity of the photonic and hadronic showers of 
energy 
above 1 
TeV   is
 $N_{\gamma}$ = 74, $N_{h}$ = 55  respectively.}
having  slightly different angles than those measured for other 
cascades 
classified as the family members. Except the hadron rich composition, this 
event has also other exotic 
feature, i.e. it is accompanied by  the surprisingly long many-maxima 
cascades (see the next subsection).

 The appearance of  families of showers at the certain apparatus layer, 
deeply 
inside the chamber, and a lack of  accompanied cascades in 
the upper part of 
the deep lead chamber ``Pamir 74/75''
 we have described also in \cite{Gladysz_Thesis}. Unfortunately, in 
these cases we were 
not able to 
exclude the trivial explanation that the observed phenomenon is simply a 
usual family, reaching the chamber during the time of its assembly.
  However, the news concerning the Centauro I remind the fact that the 
family no. 2 ($N_{\gamma}$ = 2, $N_{h}$ = 9, $\Sigma E_{\gamma}$ = 5.8 
TeV, $\Sigma {E^{\gamma}}_{h}$ = 44.3 TeV) has been accompanied by the
other six cascades (three photon and three hadron showers) with arriving 
angles only a little different
 than average $\langle \theta \rangle$ and $\langle \phi \rangle$  
measured for the main family.

{\it Could be these observations some signs of the existence of the huge 
$p_{T}$ particles?}    
 
\subsection{Strongly penetrating component}

In \cite{Gladysz_review} we emphasized the very important and up to 
now rather weakly known  
experimental aspect of the cosmic ray exotic phenomena, i.e. relation 
between Centauro species and the {\em long-flying (strongly 
penetrating) component}. In fact, the strongly penetrating component
has two faces. At first, it has been observed in the apparatus in the 
form of strongly penetrating cascades, clusters or the so called 
 halo, frequently accompanying the hadron-rich events. This 
phenomenon manifests itself by the characteristic energy 
deposition pattern revealed in shower development in the calorimeters 
indicating the slow attenuation and many maxima structure in 
homogenous thick lead chambers. The 
second face is connected 
with the strong penetrability of some objects in the passage 
through the atmosphere. Both aspects can be connected one to 
 each  other and could be the different manifestation of the 
same phenomena.

  MBHs in cosmic rays can be 
produced by cosmic neutrinos colliding with atmospheric particles 
\cite{Cavaglia_rev}. The authors of \cite{Mironov} claim that
  ``a possible neutrino-MBH origin of Centauro like events would 
explain their deep-penetration property, i.e. the fact that events 
are observed deep in the atmosphere close to the detectors.'' In 
that 
case ``there should not be any dependence of the probability of 
creating MBHs on the altitude above sea level.'' The experimental data
seem to contradict this statement. The non-observation of Centauro 
like events at Mt. Fuji (3750 m a.s.l.) and the observed smaller 
intensity  of such events at Mts Pamir ($\sim$ 4300 m a.s.l.) than 
at Mt Chacaltaya (5200 m a.s.l.)  indicate the dependence of their 
flux on the altitude in the atmosphere. Contrary to the expectation 
from the neutrino origin MBHs, if 
Centauro species were born in nucleus-nucleus collisions or if 
they were the strongly penetrating objects, produced at the top 
of the atmosphere or somewhere in the extra-galactic region, then 
the decrease of their flux with the atmosphere depth is quite 
plausible. According to \cite{Wlodarczyk_3} the estimated flux ratio 
of Centauro events $N_{Pamir}/N_{Chacaltaya} \simeq 0.07$, what 
agrees with experimental observations. 

\begin{figure}[]
\begin{center}
\mbox{
\epsfxsize=15cm
\epsfbox[-112 188 819 650]{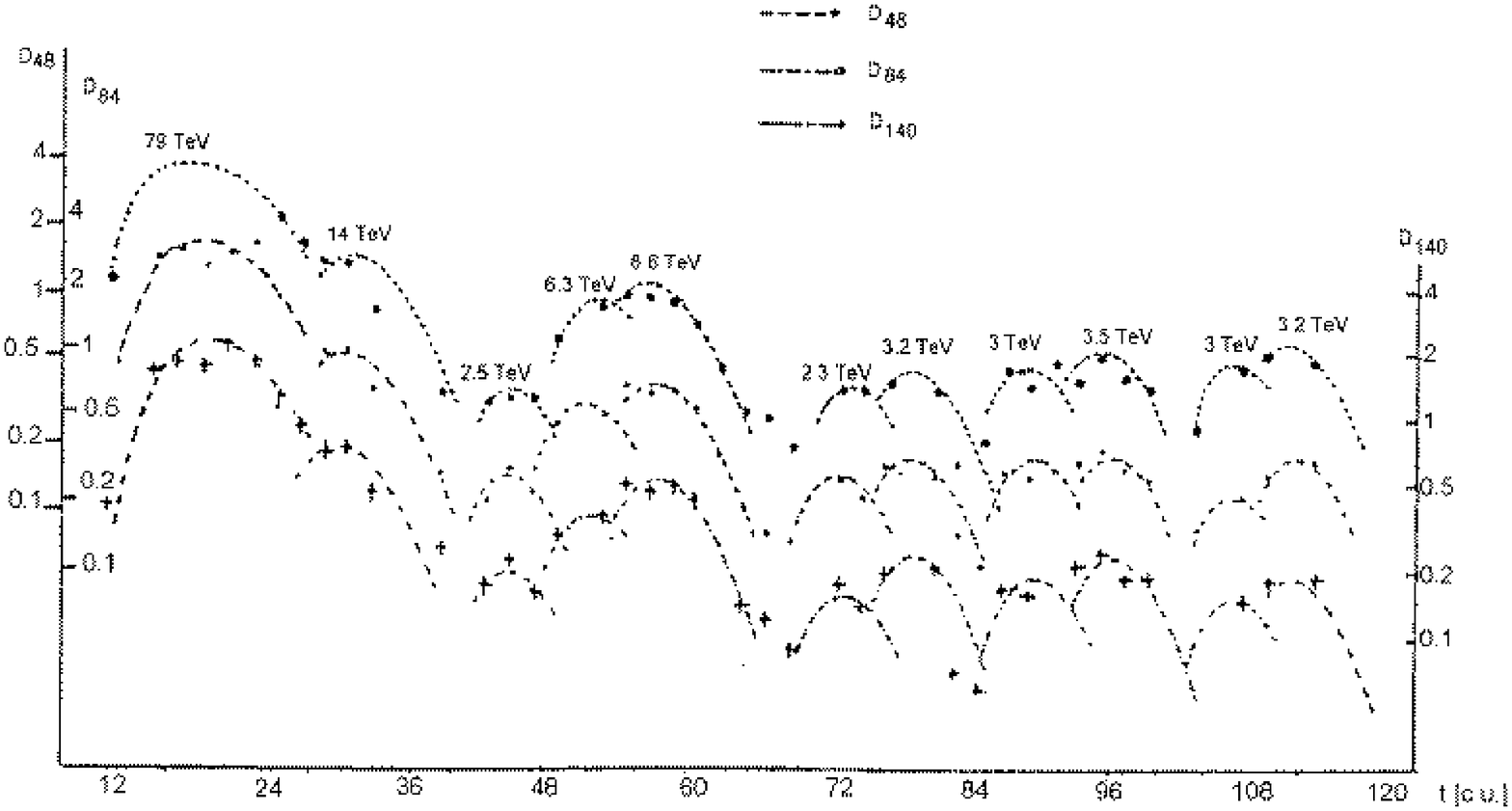}}
\hspace*{1cm}\caption[{\scriptsize Transition curves
 for cascade
no. 197.08 from Centauro C-K.}]
{\scriptsize Transition curves in X-ray film darkness D 
(measured
in
three diaphragms of a radius R = 48, 84 and 140 $\mu$) for 
cascade
no. 197.08. Energy (in TeV units) liberated  into the
soft component is indicated at each hump (averaged over three 
estimated
values) \cite{Cen_Krakow}.}
\label{fig:197}
\end{center}
\vspace*{-0.4cm}
\end{figure}

 The other serious problem of the MBH scenario seems to be 
connected with its ability to explain the second form of the strongly 
penetrating component, i.e. the unexpectedly long showers observed 
in the emulsion chambers. They are  frequently accompanying the 
Chiron and Centauro-type events. Fig.~\ref{fig:197}  
shows the 
example of such cascade found among many other hadron cascades, 
produced in the Centauro-like event \cite{Cen_Krakow} and detected in 
the 
multilayer homogenous thick Pb chamber (60 cm Pb) of the Pamir 
experiment. This exotic cascade passed a very thick layer of lead
 (109  $X_{0}$) and escaped through the bottom of the chamber.
 Its transition curve is very exotic: 11 maxima satsfactorily 
described by individual electromagnetic cascade curves can be 
revealed.

The question is {\it if and how the MBH scenario could explain the slow 
attenuation and many-maxima structure of such cascades?} It is 
true that MBHs 
are expected to produce an excited stringy states which 
according to \cite{Mironov} could play the role of the projectiles
with unusual behaviour and potentially  be responsible for the 
so-called halo events. However, such statement should be 
 explained more precisely. At the moment,  it is unclear, even 
qualitatively,  what 
 would be the nature of 
these exotic remnants of the MBH decay, and in particular which their 
features could be responsible for  the strong penetrating 
power and the 
hump structure?

  Contrary to the scenario based on MBH, 
 the main advantage of the models based on the idea of the SQM  
 is their  potential to understand the connection between  
the observation of hadron-rich events and the strongly penetrating 
component, in both above mentioned forms.
The models proposed in \cite{SQM_model_1, SQM_model_2,Shaulov,
Wlodarczyk_str} 
explain the deep penetration in the atmosphere, the model 
\cite{SQM_model_1,SQM_model_2} extended in \cite{str_cosmic} additionally 
describes 
the strongly penetrating component seen in the apparatus.  
 We are able to understand  not only Centauros but  also 
numerous hadron-rich families accompanied by 
highly
penetrating cascades, clusters or halo  by
assuming the
same mechanism of the formation of a strange quark matter fireball 
and its
successive decay into predominantly baryons and strangelet(s).

The strong penetrative power of the Centauro-like objects can be 
connected both with the small interaction cross section (in 
comparison with that of a nucleus of comparable A) of a strangelet and/or 
with big concentration of its energy in a narrow region of 
phase space. 
 This energy could be liberated into conventional particle
production in many consecutive evaporation  (or interaction) acts.

In the model described in \cite{SQM_model_1,SQM_model_2} strangelets are 
formed in 
the process of  strangeness distillation at the last stage of the 
Centauro fireball evolution. 
It is important to stress the following points:
\begin{enumerate}
\item  Characteristic features of the quark-matter nuggets, 
coming 
from the Centauro fireball decay,  have been 
estimated on the basis of experimentally measured Centauro 
characteristics.

 In particular, the strange quark density 
 and thus a strangelet mass 
number 
$A_{str}$  have been calculated from the formulas:

\begin{equation}
A_{str} \simeq N_{s\overline s} = \rho_{\overline s}V
\end{equation}
where $\rho_{\overline s}$ is $s$--antiquark density and depends on 
the
temperature by the relation \cite{Biro}:
\begin{equation}
\rho_{\overline s} \simeq 0.178(\frac{T}{200MeV})K_{2}(\frac{150
MeV}{T}).
\end{equation}
For T=130 MeV, which is the temperature for cosmic-ray
Centauros,
 $\rho_{\overline{s}}
\simeq 0.14$ fm$^{-3}$.
Evaluating  a volume of the Centauro fireball V $\sim$ O(100 
fm$^{3}$)
the number of created $s\overline{s}$ pairs and thus  
$A_{str}$
is $\sim$ 14.

\item  The indicated above value of a  strangelet mass 
number $A_{str} 
\simeq 14$ is consistent with that we
 estimated independently by using the other method and other piece 
of experimental information (analysis of the shape of the 
strongly penetrating many-maxima cascades \cite{str_cosmic}).

\item  The long many-maxima cascades observed in the thick 
lead emulsion chambers \cite{Cen_Krakow,Arisawa} have been 
satisfactorily described by the strangelet scenario \cite{str_cosmic}.

Commenting the statement from \cite{Mironov}, ``The models involving 
strangelets ... 
depend, in addition, on {\em ad hoc} assumptions about properties of 
hypothetical strange matter (both its formation and decay)'', 
 we would 
like to explain the following points:
\end{enumerate}
\begin{itemize}
\item {\em The signal produced in the thick emulsion 
chamber/calorimeter
    by the strangelet passage through the apparatus does not depend 
 on the mechanism of a strangelet formation.}

 It does not matter, if 
the strangelet is 
produced via strangeness distillation, coalescence mechanism, or 
in other quite different process. The single strangelets born by 
any 
mechanism will give the same signals in the detector.

 It is true that different 
patterns can be obtained  if the strangelet formation is 
 accompanied by production of other species. However, the 
conclusion from our 
studies is that  both the  strangelets born among other 
conventionally produced particles and the strangelets produced via 
the Centauro fireball decay give the signals quite different from 
usual events and resembling the experimentally found cascades, 
strongly 
penetrating through the apparatus.   In the papers 
\cite{Gladysz_review,str_LHC} we have shown the simulated signals 
produced by 
single 
strangelets as well as  by strangelets born among other 
conventionaly produced 
 particles, during their passage through the deep tungsten calorimeter.
  In  \cite{str_mixed}  the 
strangelets being the 
remnants of  the Centauro fireball explosion have been 
investigated. In this case the resulting signal is the sum of 
the strangelet, other Centauro decay products and the conventional 
background. In each of the considered cases the resulting 
signal is very characteristic and it can be easily distinguished from 
the conventional background. We have done calculations at energies 
predicted for $PbPb$ collisions at LHC collider  
and for the CASTOR calorimeter \cite{CASTOR}, being the subsystem of the 
CMS 
experiment. The 
upper picture at the  Fig.~\ref{fig:fig_str_mixed} shows the example
of the simulated transition curves produced 
 in the calorimeter by the  stable strangelet ($A_{str}$ = 20, 
$E_{str}$ 
= 20 TeV) (full line),
 the other Centauro decay products
 (dashed line), and the HIJING estimated background of conventionally 
produced particles 
(hatched area). 
 The lower picture shows the summed signal of these three 
contributions (full line) in comparison with the average signal  
 in the other (not being hit by the strangelet) calorimeter sectors 
(dashed line) and also 
with the predictions of the HIJING model (histogram).
\item {\em The strong penetrating power through the apparatus is 
the feature of long-lived as well as short-lived strangelets.}

\begin{figure}[]
\begin{center}

\vspace*{-5cm}

\mbox{}
\epsfig{file = 
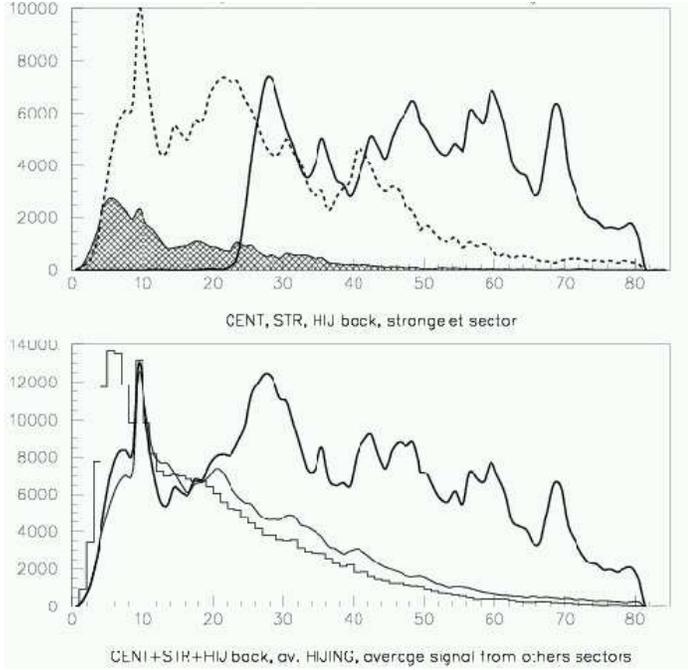,bbllx=0,bblly=0,bburx=542,bbury=648,width=10cm,clip}
\caption{{\scriptsize Transition curves produced by the Centauro and 
the strangelet being its remnant, 
 during their  passage 
through the deep calorimeter (proposed for the CASTOR detector).
 The energy deposits in the 
calorimeter versus its depth expressed in the number of 
the calorimeter 
layers (one calorimeter layer $\simeq$ 1.8 (3.7) $X_{0}$ in 
electromagnetic (hadronic) part of the calorimeter)  are shown in the 
sector hit by the strangelet.}}
\label{fig:fig_str_mixed}
\end{center}
\end{figure}

 We found in our simulations \cite{str_cosmic} that 
the long many-maxima cascades can be produced in  thick emulsion 
chambers independently  of the strangelet lifetimes.  We 
expect also that the 
nature of the baryons being the strangelet decay products weakly 
influence the shape of the produced showers.
In Fig.~\ref{fig:str_pass}  three examples of many-maxima long 
range 
cascades, obtained in simulations of the passage of different kinds of 
strangelets through the emulsion chamber, are shown.
\begin{figure}[]
\vspace*{1mm} 
\begin{minipage}{7cm}
\hbox{
\epsfxsize=180pt
\epsfbox[-211 151 690 695]{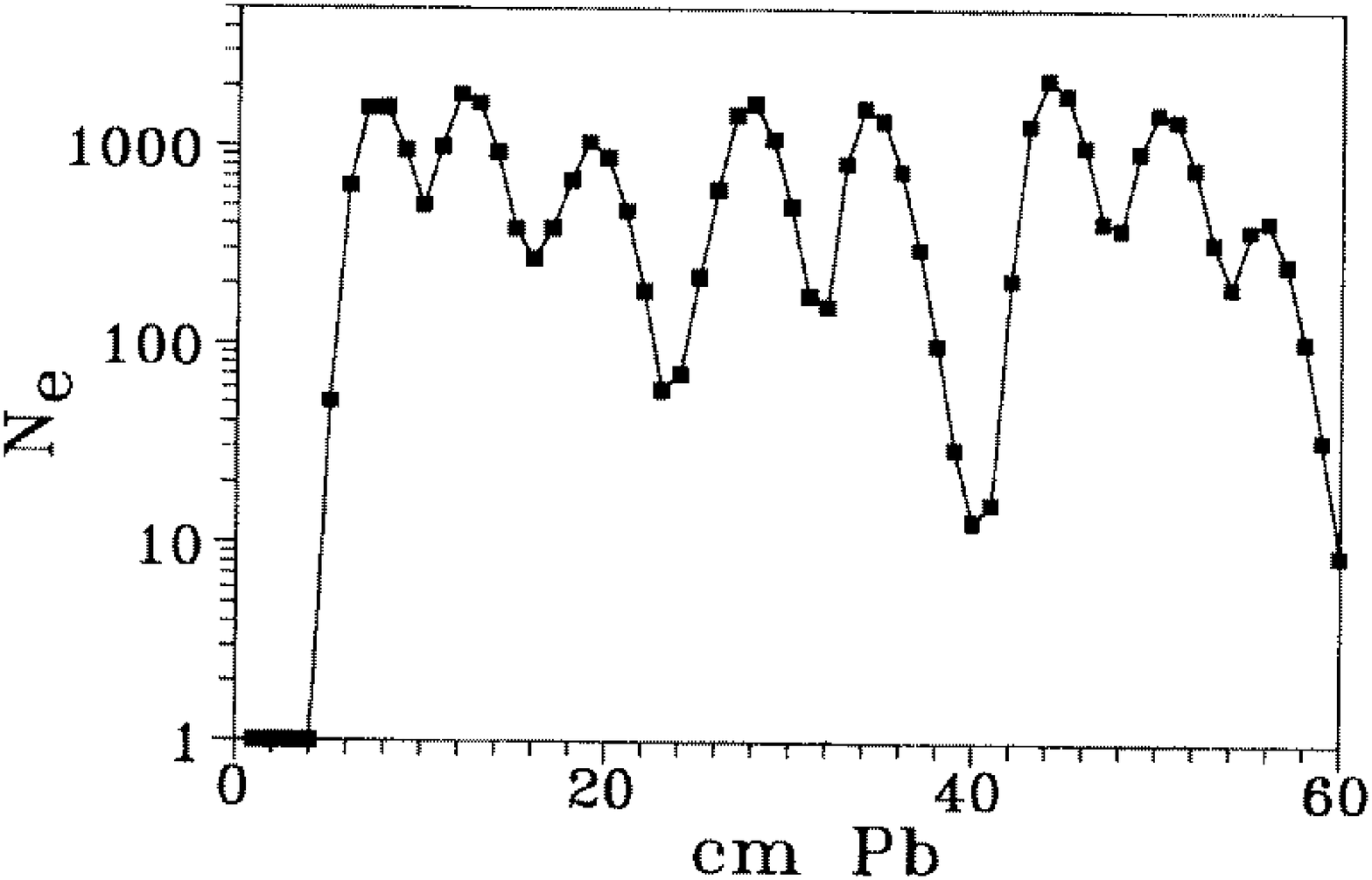}}
 \begin{center} {\scriptsize Unstable
 strangelet decaying into a bundle
 \vspace*{-1mm}\\ of 7 neutrons  
 ($E_{n} \simeq  E_{str}/A_{str} \simeq
$
 200 TeV)}.
\end{center} 
\end{minipage}
\begin{minipage}{8cm}
\vspace*{-4mm}  
\hbox{
\epsfxsize=180pt
\epsfbox[-211 97 705 
709]{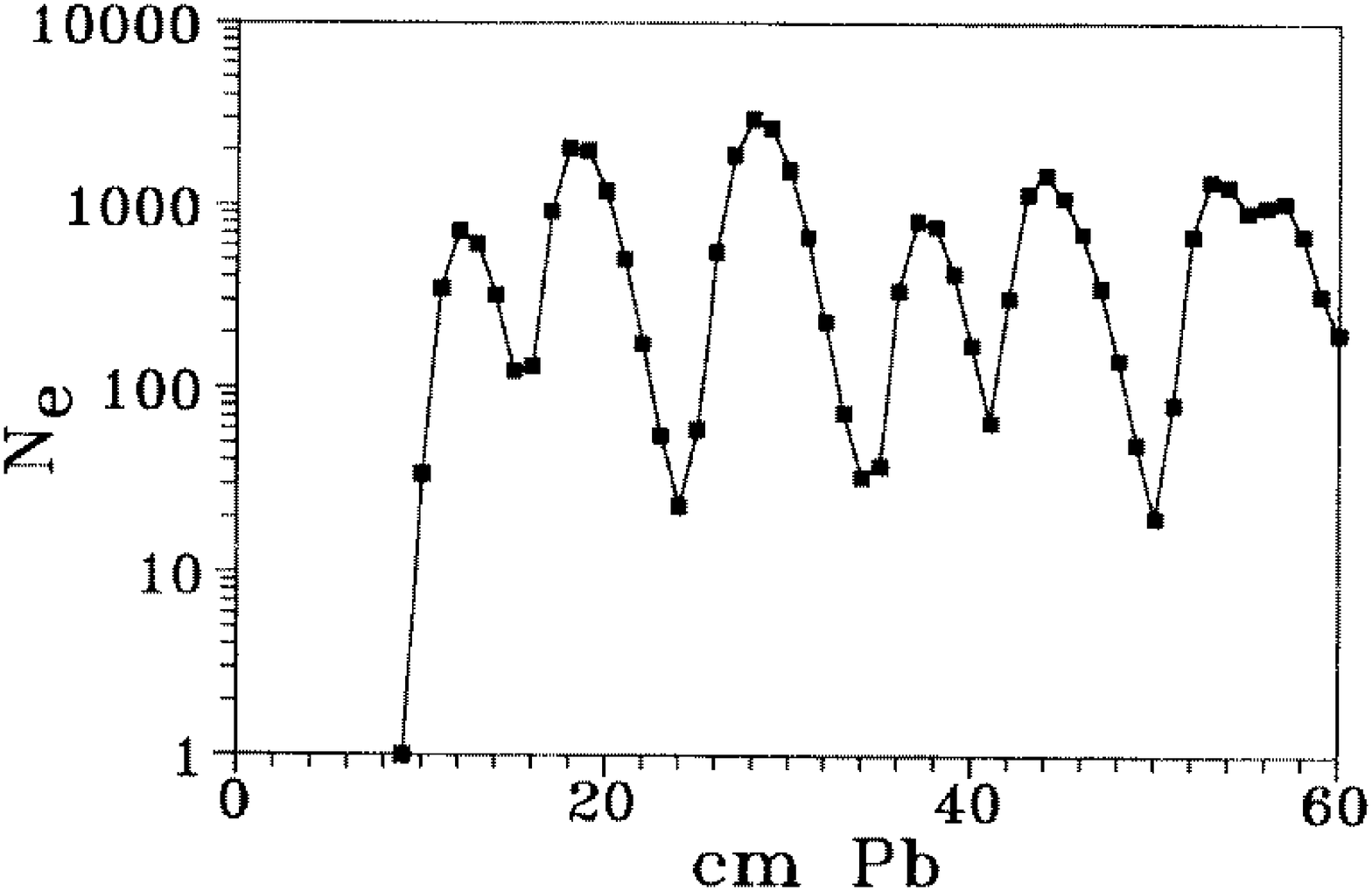}}
\vspace*{-1mm}
\begin{center} {\scriptsize Metastable strangelet\vspace*{-1mm}\\
  $(A_{str}=15, E_{str}=200$
 A TeV, $\tau \sim 10^{-15}$ s).}
\end{center}
\end{minipage}
\begin{minipage}{7cm}
\vspace*{3mm}
\hbox{
\vspace*{4mm}
\epsfxsize=180pt
\epsfbox[-211 146 795
760]{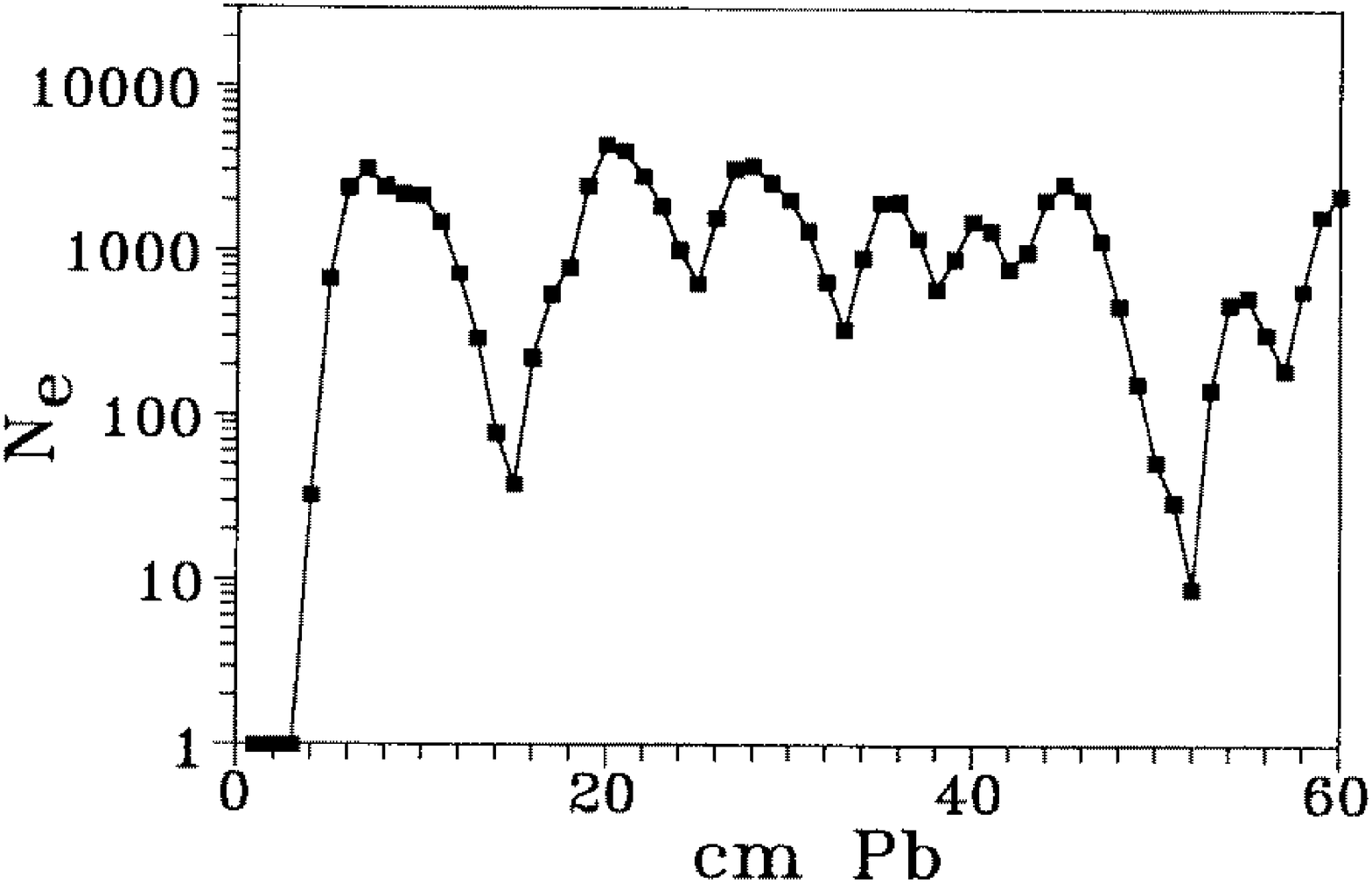}}
\begin{center}
\vspace*{-2mm}
\scriptsize{
 Long-lived strangelet\\
 ( $ A_{str} = 15$, $\mu_{q} =$ 600 MeV).}
\end{center}
\end{minipage}
\vspace*{-0.8cm}
\begin{minipage}{6cm}
\caption[{\scriptsize Examples of simulated transition curves 
recorded
in the lead chamber and produced by various strangelets.}]
{{\scriptsize Examples of simulated transition curves recorded
in the lead chamber and produced by various strangelets. Numbers of
electrons $N_{e}$ are counted within the
radius of
 50 $\mu$m.}}
\label{fig:str_pass}  
\end{minipage}
\vspace{5mm}
\end{figure}

 The thickness of 
the emulsion chambers  used in the cosmic ray mountain 
experiments is too small to distinguish between the stable 
and unstable strangelet decay scenario. On the other hand, our 
simulations done for 
the CASTOR detector \cite{CASTOR} allow to hope that  
future collider experiments, using much deeper calorimeters, 
 will be able to distinguish between 
different strangelet lifetimes.

 It is true that  predictions concerning the decay modes of such 
hypothetical objects as strangelets can not be precise and in some 
 points they are simply more or less justified speculations. The same 
remark concerns of course other exotic objects, among them  the black 
holes. At the moment  a weak dependence 
of the form of the strongly penetrating component on the 
mechanism of a strangelet formation and  its 
decay modes  strongly supports the SQM Centauro model.

\item {\em The strong penetrability power in the matter  
 is mainly the result of the big concentration of  energy
in a small range of the phase space. The strangelets as the 
dense and cold objects
with relatively high mass numbers are reasonable candidates for such 
species. The small geometrical  cross section is the additional 
factor increasing the penetrability of the long-lived 
strangelets through the matter.}

 In \cite{str_cosmic,str_LHC} we 
have calculated  the mean interaction  
 path of the stable strangelet, basing on  \cite{Fahri}.
The strangelet was considered as an object with the radius
\[R = r_{0} A^{1/3}_{str}\]
where the rescaled radius
\begin{equation}
r_{0} = ({\frac{3 \pi}{2(1 - \frac{2 \alpha_{c}}{\pi})[\mu^{3} +
(\mu^{2} - m^{2})^{3/2}]}})^{1/3}
\end{equation}
$\mu$ and $m$ are the chemical potential and the mass of the strange 
quark
respectively and $\alpha_{c}$ is the QCD coupling constant.
 
 The mean
interaction path of strangelets in the absorber (e.g. tungsten)
\begin{equation}
\lambda_{s-W} = \frac{A_{W} \cdot m_{N}}{\pi(1.12 A^{1/3}_{W} +
 r_{0}A^{1/3}_{str})^{2}}.
\end{equation}
It is clear  that the main reason of the long 
interaction lenght of a strangelet is its small geometrical radius, 
being the funcion of the  quark chemical potential $\mu$. In principle,
 $\mu$ is 
the parameter of the
model, however, two points should be mentioned. Firstly, 
the value of this parameter ($\mu$ = 600 MeV) has been estimated 
experimentally from 
the observed 
characteristics of cosmic ray Centauros. Morever, we have done 
simulations for very wide range  of  values of this parameter 
($\mu$ = 300 and 600 MeV in \cite{str_cosmic}, $\mu$ = 600 and 1000 
MeV in \cite{str_LHC}) and we found that   the interaction 
length of the strangelet is always longer than that for the nucleus 
of 
comparable mass number A, what results in the  unexpectively long 
penetrating showers. The examples of the values of the $r_{0}$ 
and $\lambda_{s-W}$ for  stable strangelet interacting in the
tungsten absorber are 
shown in Table 2.
\begin{table}[]
\label{tab:str_lambda}  
\caption{Stable strangelets interacting in the tungsten absorber}
\begin{center}
\vspace{0.3cm}
\begin{tabular}{|cccccc|}
\hline
 & & & & & \\
 $\alpha$& $A_{tg}$ & $A_{str}$ & $\mu$ [MeV]& $\lambda$ [cm] & 
$r_{0}^{str}$ \\
 &&&&&\\
\hline
\hline
&&&&&\\
0.3& 172& 15& 300& 7.3& 1.0\\
   &    & 15& 600& 10.1& 0.48\\
   &    & 20& 600& 9.7& 0.48\\
   &    & 40& 600& 8.9& 0.48\\
   &    & 20& 1000& 11.4& 0.28\\
   &    & 40&1000& 10.7&0.28\\
 &&&&&\\
\hline
\end{tabular}
\end{center} 
\end{table}

  Fig.~\ref{fig:stable_3}  shows examples of different stable 
strangelets, characterized by the values of parameters expected  
for the LHC energies and conditions (energy $E_{str}$ = 10-40 
TeV,
baryon number $A_{str}$ = 15-40, quark chemical potential $\mu_{q}$ = 
600,
1000 MeV). It is seen that all of them are 
characterized by very long attenuation pattern and they can be 
easily distinguished from the conventional events.

\end{itemize}

\section{Conclusions}
The QGP strangelet mechanism proposed in some of our papers 
offers a good  description of anomalous cosmic ray events.
In contrary, at this stage of development, the idea of explanation
 of Centauro like events via a  
decay of mini black holes 
encounters various difficulties when compared with experimental 
observations. However, some consequences of mini black hole scenario, 
such as  appearance  of huge $p_{T}$ particles, makes this idea worth 
of further considerations and  encourages for more detailed studies.

\begin{figure}[]
\begin{center}

\vspace*{-3cm}

\mbox{
\epsfxsize=280pt
\epsfbox[1 0 511 566]{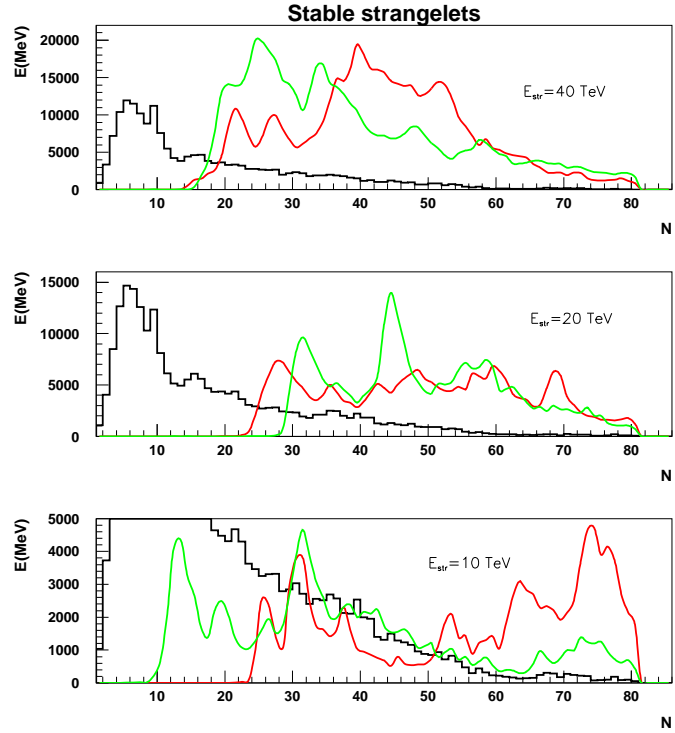}}
\end{center}
\vspace*{-1cm}
\caption [\scriptsize{
 Transition curves of stable strangelets.}]
{\scriptsize{
Transition curves of stable strangelets with energies $E_{str}$ = 
10-40 
TeV,
baryon number $A_{str}$ = 15-40, quark chemical potential $\mu_{q}$ = 
600 MeV (the red lines) and
1000 MeV (the green lines). Energy deposit (MeV) in\\ the calorimeter 
layers, in the
octant
containing a strangelet, is shown. Full line histograms show the
HIJING
estimated background.}}
\label{fig:stable_3}
\vspace*{-2mm}
\end{figure}

\vspace*{0.5cm}
{\bf Acknowledgments}

I would like to thank prof. J. Bartke for his remarks and corrections done 
 after reading of the manuscript. Thank you dr M. Cavaglia for his 
suggestions and dr 
V. Kopenkin for his critical remarks.  
This work was partly supported by Polish State Committee for Scientific 
Research grant No. 2P03B 057 24 and 
SPUB-620/E-77/SPB/CERN/P-03/DWM51/2004-2006.

\end{document}